\documentclass{article}
\usepackage{graphicx} 
\usepackage{hyperref}
\usepackage{url}

\usepackage{tikz}
\usetikzlibrary{arrows.meta, positioning}
\usepackage{msc}

\usepackage{todonotes}

\title{The Postman: A Journey of Ethical Hacking in PosteID/SPID Borderland}
\author{Gabriele Costa}
\date{IMT School for Advanced Studies Lucca}

\begin{document}

\maketitle

\begin{abstract}
This paper presents a vulnerability assessment activity that we carried out on \emph{PosteID}, the implementation of the Italian Public Digital Identity System (SPID) by Poste Italiane.
The activity led to the discovery of a critical privilege escalation vulnerability, which was eventually patched.
The overall analysis and disclosure process represents a valuable case study for the community of ethical hackers.
In this work, we present both the technical steps and the details of the disclosure process.
\end{abstract}

\section{Introduction}

Digital identity management is a cornerstone for the secure and reliable access to the services of the public administration.
In Italy, two official authentication systems exist, i.e., \emph{SPID}\footnote{\url{https://www.spid.gov.it/en/}} and \emph{CIe}\footnote{\url{https://www.cartaidentita.interno.gov.it/en/}}.
The Public Digital Identity System (SPID) is based on an open protocol specification given by the Agency for Digital Italy (AgID)\footnote{\url{https://www.agid.gov.it/sites/default/files/repository_files/spid-regole_tecniche_v1.1_0.pdf}}
All in all, the SPID authentication protocol is based on the well-known SAML Single-Sign On authentication flow \cite{samlsso}.
However, the specification was extended to also cover other aspects.
In particular, the protocol is agnostic w.r.t. the \emph{strong authentication} mechanisms that different identity providers may want to use in their implementations.
Hence, the specification include parts dedicated to which authentication factors and communication channels should be utilized.

In this work, we present a security review that, stating from the original specification and existing guidelines, allowed us to identify and demonstrate a severe privilege escalation vulnerability.
The vulnerability used to affect one of the major SPID implementations, i.e., PosteID by Poste Italiane.
Thanks to the collaboration and prompt reaction of the identity provider, the vulnerability was disclosed and patched, thus removing a serious threat for many citizens.
Our work highlighted several crucial aspects that, we advocate, are important to help the stakeholders to improve their security posture.
For this reason, together with the technical operations, here we also present the responsible disclosure process as well as some lesson learned.

\section{Timeline, engagement and disclosure}

\begin{figure}
    \centering
    \includegraphics[width=0.96\linewidth]{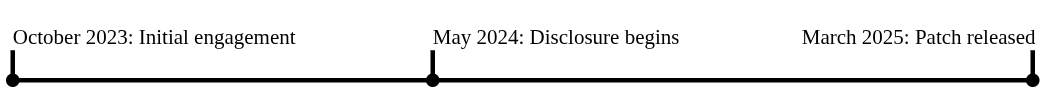}
    \caption{Timeline of the analysis activity and incident management.}
    \label{fig:timeline}
\end{figure}

The activity formally started in October 2023, when we requested the authorization to carry out a security assessment of the Android application PosteID.\footnote{\url{https://play.google.com/store/search?q=posteid}}
The vulnerabilities search was entirely focused on the communication protocol between the PosteID client App and the remote server.
Hence, we informed the staff of Poste Italiane about the type and nature of the interactions that we might have established with their remote infrastructure, including information about the real account to be used for our experiments.
Since all of the activities respected the nominal behavior of the service, they agreed with the proposed work plan.

The security assessment was conducted till May 2024, when a vulnerability was identified and confirmed through a proof-of-concept exploit.
The disclosure initiated immediately with a first communication and then continued though a series of encounters.
During these meetings the vulnerability was presented, documented and demonstrated.
As part of the disclosure, we proposed a CVSS\footnote{\url{https://www.first.org/cvss/v4-0/}} risk score of 8.7 that, eventually, was reduced to 8.3.
Also, we suggested two countermeasures to $(i)$ immediately disable the attack (by limiting some of the service functionalities) and $(ii)$ definitely fixing the vulnerability.

After the disclosure phase, Poste Italiane carried out internal processes that we could not directly observe and, in March 2025, definitively patched the PosteID app by releasing a new version.

\section{Preliminary assessment}

The preliminary phase of the security analysis consisted of the following three steps.

\paragraph{Target identification.}
In this phase we had to identify which direction was more promising for finding a vulnerability.
Clearly, code vulnerabilities were considered, but eventually we opted for \emph{flaws in the protocol design}.
The reasons behind this choice are manifold.
First, we assumed that Poste Italiane has a structured code development and review process, able to identify and fix code bugs.
Moreover, we expected (see below) strong code protection mechanisms to be in place, thus countering code analysis methods (e.g., decompilation).
Instead, we identified the authentication protocols as interesting since they are well-documented and succinct.
Also, since protocol flaws are due to design errors that are difficult to identify in their implementation, we expected possible blind spot in the code review process.

\paragraph{Information gathering.}
We revised the literature about good and bad practices in the design and implementation of strong authentication protocols. 
In particular, in~\cite{SINIGAGLIA2020101745} the authors reported that ``both NIST~\cite{NIST2017} and PCI-CSS~\cite{PCI2017} [...] deprecate the usage of out-of-band authentication via SMS''.
The reason is that, when the protocol client is executed on a mobile device, e.g., as in the case of an app, the \emph{Attacker-in-the-Device} (AidD) \cite{lapolla13} becomes overly powerful.
In other terms, the mobile device is a security bottleneck for strong authentication protocols using multiple channels that are not truly independent.

\paragraph{Technology assessment.}
The last step was related to the identification and evaluation of the technologies we had to deal with.
In general, we considered the Android OS and its Java-based app ecosystem.
Apps are distributed through zip-compressed packages called APK.\footnote{\url{https://source.android.com/docs/security}}
An APK contains various, including the application bytecode, which is provided as a single\footnote{Actually, the file can be split if it exceeds a certain size, but this has no effect on our presentation.} file called \emph{classes.dex}.
The bytecode stored there is typically generated through the compilation of Java sources and it is meant to be executed by the Android Java virtual machine, i.e., the ART VM.
Since the Android/Java bytecode is an intermediate language, \emph{decompilation}, i.e., reconstruction of the source code from the bytecode, is doable and several tools exist for that (see~\cite{MautheDec} for a survey).
To avoid that, developers typically resort to \emph{obfuscators}, i.e., tools that, without changing its behavior, scramble the code and remove any useful information such as comments and variable names.
As a result, obfuscated bytecode can still be decompiled, but the output source code is scarcely intelligible.

Beyond the programming framework, we were expecting a few more technological choices.
For instance, we expected the developers to be aware about best practices in terms of security APIs, e.g., do not re-implement cryptographic primitives.
Furthermore, due to the SPID protocol specification, we expected most of the relevant network operations to be carried out via HTTP and with state-of-the-art PKI solutions.

\section{Technical analysis and experiments}

Under the previous assumptions, we decided to proceed with the dissection anf analysis of the app. 
Our goal was to $(i)$ reconstruct the interaction between the client app and the PosteID server, and $(ii)$ analyze and hijack the strong authentication protocol implemented by PosteID.

\paragraph{Static analysis.}

Due to the industrial-level code obfuscation, our static analysis could only focus on instructions that cannot be modified, i.e., API calls.
Clearly obfuscators cannot change API names or the Java VM would be unable to handle the invocations.
Among others, by inspecting the PosteID code we identified several APIs belonging to the following packages.

\begin{itemize}
\item \textbf{javax.crypto}: a standard Java cryptography library. 
Used to generate asymmetric encryption keys.  

\item \textbf{javax.net}: a standard Java networking library. 
Used for various network-related activities, including creation and usage of SSL connections.

\item \textbf{java.security}: another standard Java security library providing encryption keys management, secure random number generation and cryptographic hash functions.

\item \textbf{org.spongycastle}:\footnote{\url{https://www.bouncycastle.org/}} a open source, multi-platform cryptographic library.
Its functionalities partially, overlaps with the standard Java APIs (see above).

\item \textbf{okttp3}:\footnote{\url{https://square.github.io/okhttp/}} a library implementing the OkHttp client, which is mainly oriented to efficiency. 
\end{itemize}

This operation was carried out by using textual searching tools like \emph{grep}\footnote{\url{https://man7.org/linux/man-pages/man1/grep.1.html}} and through manual code inspection.
Clearly, due to code obfuscation, apart from the list of API names and some sporadic pieces of data, we could not infer useful details, e.g., about the API parameters and return values.

\paragraph{Dynamic analysis.}
The next goal was to reconstruct the exact interaction between the client app and PosteID server.
This basically reduced to the following two problems:
\begin{enumerate}
    \item observe the data exchanged between the two parties, and;
    \item eliminate all the irrelevant app functionalities.
\end{enumerate}
To obtain this result, we opted for \emph{dynamic slicing}~\cite{slicing,slicingandroid}.
Briefly, this technique generates a subset of the instructions executed by a program at runtime.
The resulting instructions, i.e., the code slice, represent a linear fragment of the original program that, under proper assumptions, has its same behavior.
The main advantage of dynamic slicing is that, when correctly configured, slices include all the real data used by the program during its execution and only the relevant instructions necessary to manipulate them.
So, for instance, in our case, a slice should contain both the API calls needed to send and receive HTTP messages and the API calls used to encrypt and decrypt these communications.
In general, the main drawback of dynamic slicing is that it can only capture a single execution of the program and its internal branching logic is completely neglected.
However, in our case, this was not a limitation as network protocols usually consist of a single execution flow, where the only conditions are related to failures (e.g., in case of malformed messages).
Furthermore, in order to generate a valid program slice, we needed to instrument the application code.
To do that, we resorted to \emph{LSPosed},\footnote{\url{https://github.com/LSPosed/LSPosed}} a framework for statically rewriting APK bytecode.
In particular, before and after each API call of interest, we added instructions for logging the API name, parameters and return value.
Finally, we installed the modified PosteID APK on a real device,\footnote{Notice that the app also includes checks on the execution platform to avoid installation on virtual devices.} we executed the application under some different scenarios, and we collected the corresponding slices.

\paragraph{Protocol hijacking.}

Program slices were then analyzed individually and tested to ensure the repeatability of all the protocol execution flows.
Furthermore, among all the protocol flows, we isolated those that include the use of the out-of-band SMS channel.
In particular, the \emph{device activation} flow was immediately identified as interesting.
The flow is executed once for every fresh installation of the PosteID application.
The goal is to enroll the user's device and register it to operate as an authorized SPID client.

\begin{figure}[t]
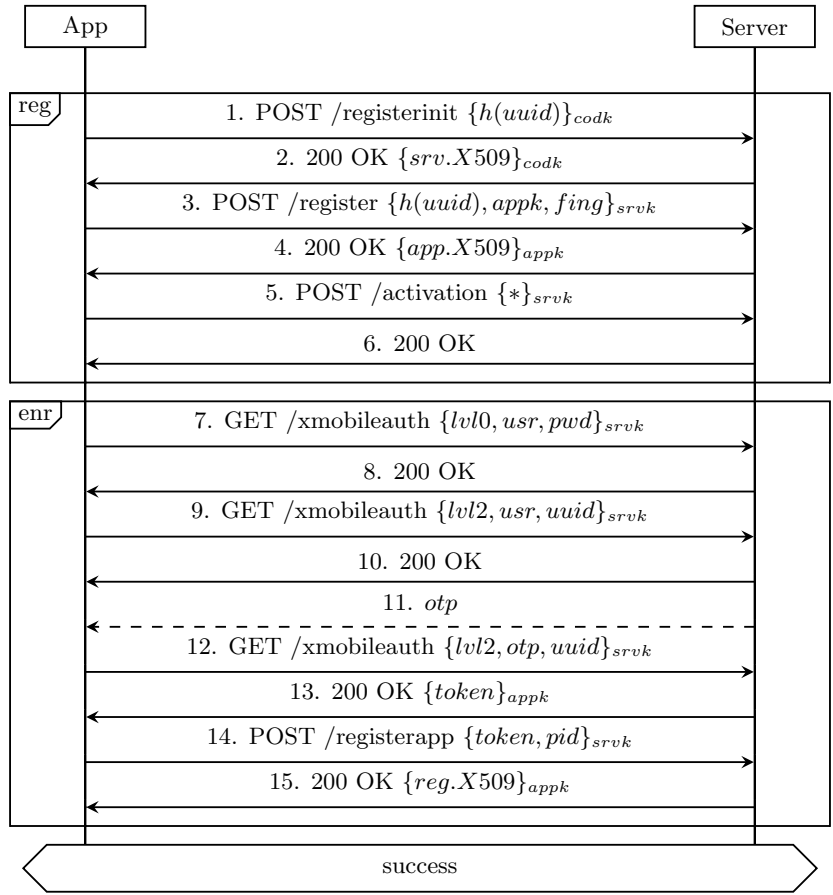

    \centering
\begin{msc}[
font=\small,
head top distance=0cm,
foot distance=0cm,
environment distance=0cm,
instance distance=0.6\columnwidth]{}
\setmsckeyword{}
\drawframe{no}
\drawinstfoot{no}
\declinst{app}{}{App}
\declinst{srv}{}{Server}

\inlinestart{reg}{reg}{app}{srv}
\nextlevel[1.2]
\mess{1. POST /registerinit $\{h(uuid)\}_{codk}$}{app}{srv}
\nextlevel[1.2]
\mess{2. 200 OK $\{srv.X509\}_{codk}$}{srv}{app}
\nextlevel[1.2]
\mess{3. POST /register $\{h(uuid), appk, fing\}_{srvk}$}{app}{srv}
\nextlevel[1.2]
\mess{4. 200 OK $\{app.X509\}_{appk}$}{srv}{app}
\nextlevel[1.2]
\mess{5. POST /activation $\{*\}_{srvk}$}{app}{srv}
\nextlevel[1.2]
\mess{6. 200 OK}{srv}{app}
\nextlevel[0.5]
\inlineend{reg}
\nextlevel[0.5]
\inlinestart{enr}{enr}{app}{srv}
\nextlevel[1.2]
\mess{7. GET /xmobileauth $\{lvl0, usr, pwd\}_{srvk}$}{app}{srv}
\nextlevel[1.2]
\mess{8. 200 OK}{srv}{app}
\nextlevel[1.2]
\mess{9. GET /xmobileauth $\{lvl2, usr, uuid\}_{srvk}$}{app}{srv}
\nextlevel[1.2]
\mess{10. 200 OK}{srv}{app}
\nextlevel[1.2]
\mess*{11. $otp$}{srv}{app}
\nextlevel[1.2]
\mess{12. GET /xmobileauth $\{lvl2, otp, uuid\}_{srvk}$}{app}{srv}
\nextlevel[1.2]
\mess{13. 200 OK $\{token\}_{appk}$}{srv}{app}
\nextlevel[1.2]
\mess{14. POST /registerapp $\{token, pid\}_{srvk}$}{app}{srv}
\nextlevel[1.2]
\mess{15. 200 OK $\{reg.X509\}_{appk}$}{srv}{app}
\nextlevel[0.5]
\inlineend{enr}
\nextlevel[0.5]
\condition{success}{app,srv}
\end{msc}
\caption{PosteID new device activation protocol.}
\label{fig:msc}
\end{figure}

The actual message sequence is depicted in Figure~\ref{fig:msc}.
Briefly, a freshly installed app automatically starts the registration phase (reg) by sending (1.) a /registerinit request that contains the hash code of a generated $uuid$\footnote{\url{https://en.wikipedia.org/wiki/Universally_unique_identifier}} identifier.
Also, the request is encrypted with a symmetric key $codk$, hard-coded in the application.
Then, the server answers (2.) by sending its X.509 certificate (still encrypted with $codk$).
From now on, the client encrypts its outgoing messages with the public key of the server $srvk$.
So, the application continues (3.) with a /register request containing the hash of its $uuid$, its own public key $appk$ and a system fingerprint $fing$.
Briefly, $fing$ is a string encoding some OS and hardware data generated by the app (e.g., the OS version and whether the device is rooted).
The server (4.) responds by sending other X.509 certificates for the application to be used in future interactions, and the registration ends (5. and 6.) with a call to the /activation endpoint.

Now the application is registered, but it is not authorized for the strong (level 2) authentication.
For that, the enrollment (enr) phase is needed.
Initially (7. and 8.) the application authenticates to /xmobileauth with basic ($lvl0$) credentials ($usr$ and $pwd$).
Then, it calls /xmobileauth again (9.) to elevate its privileges to $lvl2$.
After confirmation (10.) the server also sends via SMS at a registered phone number (11. dashed arrow) a one-time password ($otp$) that the application has to resubmit (12.) in order to demonstrate that the user controls the registered SIM card.
If the correct $otp$ is received, the server answers (13.) with an authentication token ($token$).
The token can be used to register a new identification code $pid$ (a.k.a. the ``PosteID code'') that the application will use for future level 2 authorization.
Finally, the server answers (15.) by returning a X.509 certificate associated with the enrolled device and the protocol successfully ends.

\section{Attack narrative and impact}

As previously stated, the attack scenario is that of an AitD, where the goal of the adversary is to obtain level 2 authentication to the account of a victim.
The capabilities of the attacker are the following.
\begin{enumerate}
    \item[A.] She can install a trojan application on the victim's device.
    \item[B.] The trojan has access to the basic credentials (level 0, i.e., username and password) of the victim.
    \item[C.] The trojan has the permissions to read incoming SMSs and use the network on the victim's device.
\end{enumerate}
Under these assumptions, the adversary performs a \emph{privilege escalation attack}.

\paragraph{Attack preparation.}
To exemplify how an attack leveraging the found vulnerability may occur in real life, we propose the following narrative.
The attacker develops a malicious application offering some services related to the ecosystem of Poste Italiane.
For instance, the malicious app could provide a map of the post offices close to the current location, news about the products of Poste Italiane or a tracker for expeditions.
The goal of the attacker is to push as many PosteID users as possible to install the trojan app (A) on a device already enabled for level 2 authentication.
Also, the provided service should not require high-level permissions, but only basic authentication (B).
In this way, the users might be reassured about the risk related to the app.
Finally, the app's service should be consistent with the permissions of using the network and accessing the incoming SMSs (C).
While using the network is reasonable for an app interacting with remote services, SMS may be justified, e.g., by the necessity of receiving off-line notifications.

\begin{figure} 
    \centering
    \includegraphics[width=0.8\linewidth]{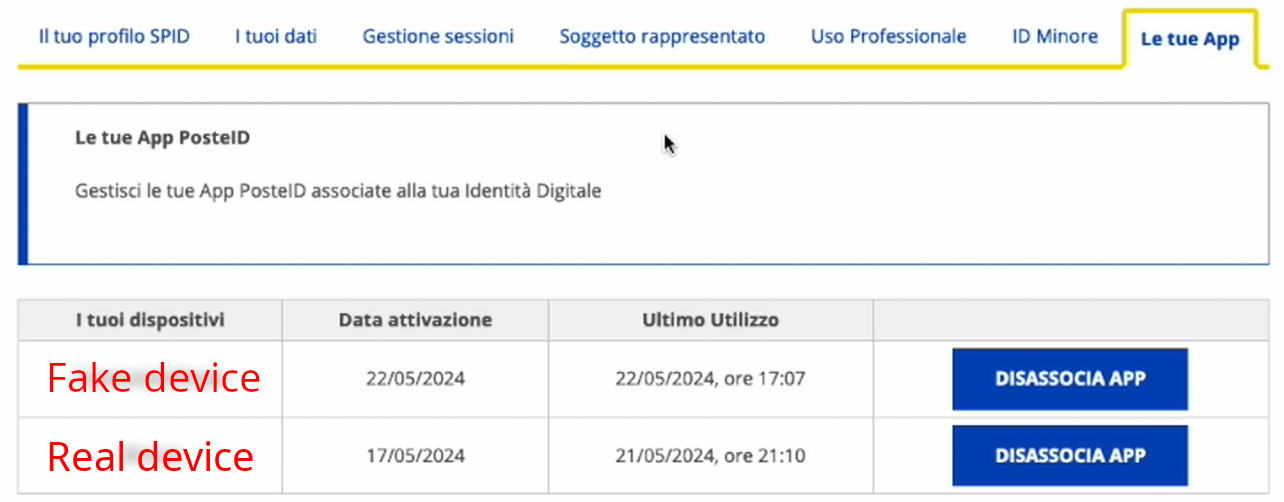}
    \caption{The app management interface with the ``disable app'' (blue) button.}
    \label{fig:dauth}
\end{figure}

\paragraph{Attack execution.}
When the trojan app has been installed on enough devices, the attacker starts the vulnerability exploitation.
In particular, through a command and control infrastructure, the attacker triggers all the installed trojan which, simultaneously, do the following.
\begin{enumerate}
    \item \textbf{Fake registration.} The trojan runs the registration phase (see Figure~\ref{fig:msc}) claiming to be a fresh installation of PosteID on a fake device.
    To do that, the trojan app needs the hardcoded key $codk$, which can be extracted from the original PosteID APK, and a system fingerprint $fing$.
    Since $fing$ is generated by combining public information, e.g., the OS version, the trojan app can algorithmically create it.
    \item \textbf{Fake enrollment.} Now the authentication server believes that the user installed two instances of PosteID on two distinct devices, the second being the fake one.
    Subsequently, the trojan app starts the enrollment phase.
    Since basic credentials are known, the only challenge is submitting the $otp$ code.
    However, the code can be intercepted and submitted by reading the incoming SMS (step 11., see Figure~\ref{fig:msc}).
    Then, the protocol can be successfully closed and the trojan app obtains the level 2 credentials.
    \item \textbf{De-registration.} Clearly, the user might notice that something weird is happening, e.g., since an unexpected SMS is received.
    Thus, the malicious app must cut the user out.
    However, level 2 credentials are sufficient for managing the registered devices and the trojan can remove the real device from the list of registered ones, only keeping itself as the only one authorized to operate for the user (see Figure~\ref{fig:dauth}).
    The user can still access her account, but since she cannot use her mobile device, she must resort to other, slower procedures such as contacting the support service.
    \item \textbf{Attack finalization.} At this point, the trojan is the only app that can impersonate the user to access critical services.
    In particular, the attacker can access private data and operate through SPID-enabled services having legal validity.
    Furthermore, if the victim has a bank account with Poste Italiane, the trojan can authorize money transfer.
\end{enumerate}

Although difficult to estimate, the potential impact of such an attack scenario appears dramatic in numbers.
Indeed, considering that Poste Italiane handles 6.5 million private bank accounts\footnote{\url{https://www.posteitaliane.it/it/performance-finanziaria.html}} and the Android version of PosteID has more than 10 million downloads,\footnote{\url{https://play.google.com/store/apps/details?id=posteitaliane.posteapp.appposteid}} even a few thousand infected devices might have resulted in a significant loss.

\section{Lesson learned}

There are several aspects related to the activity described above that deserve to be considered.

\paragraph{Role of guidelines and best practices.}
All in all, the discovered vulnerability is the result of having overlooked the existing guidelines about the adoption of SMS in strong authentication protocols.
This outlines the fundamental role of tasks that are far from the actual code writing and testing which, yet, do have a profound and practical impact. 
These tasks include the review of existing guidelines, the study of best practices and standards, threat modeling and more.

\paragraph{Design flaws implications.}
When something goes wrong with the early design of a software product, as in this case, defining and implementing an effective remediation plan may not be easy.
In fact, on top of the flawed design, many implementation choices have been taken and fixing the issue could require rethinking the entire system.

\paragraph{Technology and protection mechanisms.}
It is fundamental to underline that many security mechanisms do not prevent attacks.
Rather, they just complicate operations that, eventually, an adversary will carry out, perhaps after selecting the more appropriate tools.
In our case, industry-level obfuscation was proven almost completely ineffective against dynamic slicing.

\paragraph{Responsible disclosure.}
The responsible disclosure process still lacks of standards and regulation.
As a consequence, the steps are the result of a negotiation between the ethical hackers and the system owner.
In practice, this slows down the entire process, makes it less reliable and complex for small and medium companies that might miss the appropriate security facilities.

\section{Conclusion}

In this paper we presented a security analysis of the PosteID app, implementing the national digital identity protocol SPID, that culminated in the detection, disclosure and correction of a dangerous privilege escalation vulnerability.
Beyond the mere technical interest, we also presented the overal process that we carried out, from the initial setup and tool selection to the eventual disclosure and lesson learned.
In general, we believe that the present work can serve to the security community, as well as the private stakeholder, to better further develop the security posture.

\section*{Acknowledgments}
The author thanks Federico Chiesa, who carried out most of the technical analysis and experiments presented in this work. 
Also, the author thanks Rocco Mammoliti and the staff of Poste Italiane that supported the anlysis and disclosure process.

This work was partially supported by project SEcurity and RIghts in the CyberSpace SERICS PE0000014 - PNRR M4C2 I.1.3, financed by the European union Next Generation EU - CUP: D67G22000340001.

\bibliographystyle{plain}
\bibliography{biblio}

\end{document}